\documentclass[pra,twocolumn,showpacs,preprintnumbers,amsmath,amssymb]{revtex4}
\usepackage[english]{babel}
\usepackage{indentfirst}
\usepackage{epsfig}
\usepackage{amsfonts,amssymb,latexsym,amsmath,enumerate}
\usepackage{bm}

\def\diag{{\rm diag}}

\begin{document}


\title{Twisted electron in a strong laser wave}
\author{Dmitry V. Karlovets}
\email{d.karlovets@gmail.com}
\affiliation{Tomsk Polytechnic University, Lenina 30, 634050, Tomsk, Russian Federation}

\date{\today}

\begin{abstract}
Electrons carrying orbital angular momentum (OAM) have recently been discovered theoretically and obtained experimentally that opens up possibilities for using them in high-energy physics. We consider such a twisted electron moving in external field of a plane electromagnetic wave and study how this field influences the electron's OAM. Being motivated by the development of high-power lasers, we focus our attention on a classically strong field regime for which $-e^2 \bar {A^2}/m_e^2 c^4 \gtrsim 1$. It is shown that along with the well-known ``plane-wave'' Volkov solution, Dirac equation also has the ``non-plane-wave'' solutions, which possess OAM and a spin-orbit coupling, and generalize the free-electron's Bessel states. Motion of the electron with OAM in a circularly polarized laser wave reveals a twofold character: the wave-packet center moves along a classical helical trajectory with some quantum transverse broadening (due to OAM) existing even for a free electron. Using the twisted states, we calculate the electron's total angular momentum and predict its shift in the strong-field regime that is analogous to the well-known shifts of the electron's momentum and mass (and to a less known shift of its spin) in intense fields. Since the electron's effective angular momentum is conserved in a plane wave, as well as in some more general field configurations, we discuss several possibilities for accelerating non-relativistic twisted electrons by using the focused and combined electromagnetic fields.
\end{abstract}





\maketitle

\section{Introduction} 

In recent years, photons carrying orbital angular momentum (OAM) have become objects of an intensive study that led to creation of a new subfield in quantum optics (see a recent review in \cite{Tw}). The beams of such \textit{twisted photons} are no longer the plane waves, since their wavefront rotates around the propagation axis, and the Poynting vector looks like a corkscrew. The simplest objects revealing these properties were shown to be the focused laser beams, in particular, the Laguerre-Gaussian wave packets \cite{Allen}. These photons have already found many different applications in condensed matter physics, atomic physics, biology, microscopy, etc. since they are used, for instance, as the optical tweezers for trapping and moving different nano- and micro-objects \cite{Fr, Grier, Liu}. It has recently been shown that rotating black holes can produce OAM in the photons radiated near them \cite{T, B}. A way of obtaining the twisted photons with the energies up to GeV range via the Compton backscattering process has recently been proposed in \cite{Serbo}.

It is evident that the massive particles, for example electrons, can also carry OAM being quantized along the propagation axis. Such non-plane-wave solutions for the Dirac equation, which describe an electron possessing some OAM along with the spin, were given, for example, in \cite{O, Bagrov, Leary}. The method for creating such electrons was proposed by Bliokh et al. in \cite{Bliokh-07} by analogy with the twisted photons, for which the spiral phase plates or diffraction gratings with an edge dislocation are often used. Soon after this, the \textit{twisted electrons} were obtained experimentally by several groups with the values of the OAM-projection $m$ onto the direction of motion up to $m \sim 100 \hbar$ \cite{M, V, Mc} (see also more recent results in \cite{S, V2}). Though the energies of these electrons are not high yet, $\varepsilon_c \sim 300$ keV, their creation itself opens up possibilities for using twisted particles in high-energy physics. The simplest quantum processes with the twisted scalar particles were studied theoretically by Ivanov and Serbo in \cite{Ivanov, I-S}. 

The OAM as a \textit{fundamentally new} quantum degree of freedom of massive particles leads to a number of intriguing effects, for instance, to some quantum broadening of the free electron's classical rectilinear trajectory together with the appearance of a ``doughnut'' spatial structure in the transverse plane analogous to that of the focused laser beams \cite{Bliokh-07}. Moreover, the gyromagnetic ratio for such twisted electrons is modified: the electron's magnetic moment becomes proportional to the OAM (with a $g$-factor equal to 1) \cite{Bliokh-11}, whose values can be, in principle, \textit{arbitrary large}. For electrons with $m \sim 100 \hbar$, the magnetic moment is, roughly speaking, $10^2$ times larger than the Bohr magneton that can lead to the different phenomena in particle and nuclear physics, for instance, to the significant increase of all the radiation effects related to the magnetic moments. Whereas the ordinary magnetic moment's radiation (the so-called ``spin light'' \cite{Bord}) is usually too tiny to be measured, the radiation of the twisted electron's magnetic moment (regardless the radiation type) is likely to be detected much easier, especially, in the coherent regime of a beam's emission (see e.g. \cite{Gover}).

However in order to make quantitative predictions on these effects, it is necessary in most cases to have the \textit{ultrarelativistic} twisted electrons (e.g. in order to increase the radiated power) as well as to understand how the electron's OAM is changed when some external electromagnetic field is applied. Since the direct experimental production of the ultrarelativistic twisted electrons seems to be impossible due to extremely small de Broglie wave length, the only way we are left with is to use some accelerating field. 

In this paper, we consider a twisted electron moving in external field of a plane electromagnetic wave and present the corresponding exact solution of the Dirac equation. The study of twisted states in the background fields is important at least for two reasons: 
\begin{itemize}
\item
it allows one to describe motion of the particles in these fields as well as the conservation (non-conservation) of the OAM that, consequently, can suggest a way for accelerating twisted electrons up to GeV-energy range; 
\item
it makes possible to bring these particles in the subject of high-energy physics and to calculate the corresponding quantum processes the simplest of which is the Compton scattering of a plane-wave photon by a twisted electron (or vice versa \cite{Serbo}). 
\end{itemize}
The investigation of these collisions can be of both the fundamental and applied interest, since the value of OAM can be arbitrary large and its importance for different quantum processes may turn out to be much greater than that of the spin. 

Thus, one of the paper's main goals is to understand whether the external laser field changes the properties of the twisted electron or not. In particular, we shall demonstrate that the average angular momentum of a twisted electron is conserved in a plane wave appearing as an effective integral of motion. In fact, our main conclusion of the OAM's conservation stays valid for even more general field configurations, since, as was demonstrated by Bagrov and Gitman, the problem with a combination of longitudinal electric and magnetic fields, $E_z + H_z$, together with the co-propagating laser field $A$ (with a wave vector ${\bf k} = k {\bf e}_z$) can always be effectively reduced to the one with the longitudinal magnetic field $H_z$ only by using the special transformation \cite{Bagrov}. As is well-known, the electron OAM's z-projection is an exact integral of motion in the last case (see e.g. \cite{Bagrov-02, Bagrov-07}) by analogy with a free twisted electron. These facts make the acceleration of the non-relativistic twisted electrons possible.

On the other hand, here we try to connect this newly developed physics of twisted particles together with the strong-field QED. The development of high-power lasers could allow one to achieve the laser intensities exceeding $10^{23}\ \text {W}\ \text{cm}^{-2}$ within the next few years (see, for example, Extreme Light Infrastructure project \cite{ELI}). When being observed in the rest frame of a relativistic electron, the field strength of such a laser can approach to the Sauter-Schwinger limit $E_c \simeq 1.3 \times 10^{18}\ \text {V}\ \text{m}^{-1}$ that leads to the different non-linear quantum phenomena (see e.g. \cite{Lim, Bulanov} and the references cited therein). Being motivated by this fact, we focus our attention on the case of the strong laser wave, which is characterized with the well-known ``classical'' parameter $\eta^2 = -e^2 \bar {A^2}/m_e^2 c^4$ (see $\S$101 in \cite{BLP}). 

The electron's twisted states presented in this paper generalize the free-electron's Bessel states recently obtained by Bliokh et al. in \cite{Bliokh-11}. They are, in addition, the simplest example of the non-Volkov solutions for the Dirac equation with the external plane-wave field, as was demonstrated in \cite{Bagrov}. Using these states, we calculate the electron's effective angular momentum (OAM + spin) in a wave and predict its shift in the strong-field regime for which $\eta \gtrsim 1$. Such a shift appears due to the electron's spin precession and it is analogous to the well-known change of the electron's momentum and mass in a strong laser wave. Though this effect is negligible when $m \gg \hbar$, it can lead to the effective reduction of spin (for helicity states: $\lambda \rightarrow 0$ for $\eta \gtrsim 1$), or even to inversion of the one ($\lambda \rightarrow - \lambda$ for $\eta \gg 1$), that can, consequently, influence some quantum processes such as the non-linear Compton scattering and the Breit-Wheeler pair production. 

Finally, we discuss several possibilities for accelerating non-relativistic twisted electrons with the OAM conservation and show that such an acceleration is possible with the use of azimuthally symmetric electromagnetic fields such as the focused laser beams and the ``combined'' fields (plane wave + longitudinal electric and magnetic fields). 

The paper is organized as follows. In Sec.\ref{Sect1} we consider a free twisted scalar particle and discuss some general moments in solving the Klein-Gordon equation and normalizing its solutions. The free twisted electron being described with the corresponding solution of the Dirac equation is considered in Sec.\ref{Sect2}. In Sec.\ref{Sect3} we study the electron moving in the field of a plane circularly polarized electromagnetic wave and present the corresponding non-Volkov states. We also calculate the electron's current density using these states and discuss the normalization. At each step we pay attention to the transformation of the twisted states to the well-known plane waves in the corresponding limiting case. Then in Sec.\ref{Sect4} we calculate the electron's effective spin and study its shift in the strong wave regime. The total (OAM + spin) electron's angular momentum in the laser wave is calculated in Sec.\ref{Sect5}. We discuss in Sec.\ref{Sect6} the possibilities for accelerating non-relativistic twisted electrons and also some new effects the OAM can bring into high-energy physics. We conclude in Sec.\ref{Sect7}. System of units $\hbar = c = 1$ and the metric with a signature $(+---)$ are used throughout the paper. 

\section{\label{Sect1} Free twisted scalar} 

Consider a quantum system described with a vector $|\psi \rangle$ that obeys the Klein-Gordon equation. Let us expand this state over the plane waves:
\begin{eqnarray}
\displaystyle |\psi \rangle = \int \frac{d^4 p}{(2 \pi)^4} \psi (p) |p\rangle \label{Eq0}
\end{eqnarray}
with $|p\rangle \propto e^{-ipr}, p^2 = m_e^2$. The projection onto the plane-wave state, $\psi (p) = \langle p|\psi\rangle$, determines the physical model the initial state $|\psi\rangle$ corresponds to. In the simplest case, the choice 
\begin{equation} 
\psi (p) = (2 \pi)^4 \delta^{(4)} (p - \tilde{p}) \nonumber
\end{equation} 
gives the ordinary plane wave $|\psi\rangle = |\tilde{p}\rangle$ usually utilized within the framework of high-energy physics. It is clear that such an expansion is possible only if the first state represents a complete set of functions that allows one to invert the expansion.

The free scalar state with a given projection of OAM onto the direction of average motion (z axis) is defined as a state with the definite energy $\varepsilon$, longitudinal momentum $p_{\parallel }$, and the absolute value of the transverse momentum $\kappa$. We shall call this the \textit{twisted state}. According to this definition, one can choose:
\begin{eqnarray}
\psi (p) = (2 \pi )^3 (- i)^m \delta (p_0 - \varepsilon) \delta (p_z - p_{\parallel }) \delta (p_{\perp }-\kappa ) \frac{e^{im\phi_p}}{p_{\perp }},
\label{Eq00}
\end{eqnarray}
where $\phi_p$ is the azimuthal angle of the momentum ${\bm p}$, and $m = 0, \pm 1, \pm 2, ...$ can be called the azimuthal quantum number \footnote{Do not confuse this with the particle's mass denoted as $m_e$!}. 
Inserting (\ref{Eq00}) into (\ref{Eq0}), we have the following state
\begin{eqnarray}
\displaystyle |\psi \rangle \equiv |\varepsilon, p_{\parallel }, \kappa, m\rangle = N J_m (\rho \kappa ) e^{-i \varepsilon t + i p_{\parallel } z + i m\phi_r}.
\label{Eq000}
\end{eqnarray}
Here, $J_m$ is the Bessel function, $\phi_r$ is the azimuthal angle of the vector ${\bm r} = \{{\bm \rho}, z\}$, and $N$ is a normalization constant. As can be seen, this state is an eigenfunction of the OAM's z-projection operator: 
\begin{eqnarray}
\displaystyle
\hat {l}_z = [\hat{{\bm r}} \times \hat {\bm p}]_z = - i \frac{\partial}{\partial \phi_r}. 
\nonumber
\end{eqnarray}
It can also be shown with the use of the Bessel functions recurrence relations that the transverse momentum of this state is $\langle \psi | \hat{p}_{\perp}| \psi \rangle/\langle \psi | \psi \rangle = \kappa$.

A transition to the plane-wave case is performed when $\kappa \rightarrow 0$, so that
\begin{eqnarray}
\displaystyle J_m (\rho \kappa ) \approx \frac{(\rho \kappa )^m}{2^m \Gamma (1 + m)} \rightarrow \delta_{m, 0},\ |\psi \rangle \rightarrow N e^{-i \varepsilon t + i p_{\parallel } z }.
\label{Eq0000}
\end{eqnarray}
Here and below one can put $m \geq 0$ without a loss of generality. In what follows, we shall call the infinitesimal $\kappa$s limiting case as the \textit{paraxial limit}.

It can be easily checked that the states (\ref{Eq000}) represent an orthogonal and complete set of functions in the infinite volume:
\begin{eqnarray}
&& \displaystyle \int d^3 r\ \psi^*_{\varepsilon, p^{\prime}_{\parallel }, \kappa^{\prime}, m^{\prime}}(r) \psi_{\varepsilon, p_{\parallel }, \kappa, m} (r) = \cr && \displaystyle = N^2 (2 \pi )^2 \delta_{m m^{\prime}} \delta ( p_{\parallel } - p_{\parallel }^{\prime}) \frac{1}{\kappa}\delta{(\kappa - \kappa^{\prime})}, \cr && \displaystyle \sum \limits_{m} \int \frac{d^3 p}{(2 \pi )^3}\ \psi^{*}_{\varepsilon, p_{\parallel }, \kappa, m} (r^{\prime}) \psi_{\varepsilon, p_{\parallel }, \kappa, m} (r) = \cr && \displaystyle = N^2 \delta ({\bm r} - {\bm r}^{\prime}).
\label{Eq0000-3}
\end{eqnarray}

The normalization constant $N$ can be chosen from the condition
\begin{eqnarray}
\displaystyle \int \limits_V d^3 r\ j^0 = 1, \ j^{\mu} = \psi^{*} i \partial^{\mu} \psi + c.c. \label{Eq0000-5}
\end{eqnarray}
where the integration is carried out over the large, but finite cylindrical volume $V = \pi R^2 L$. Taking into account Eq.(\ref{Eq000}), we have
\begin{eqnarray}
&& \displaystyle \int \limits_V d^3 r\ j^0 = 2\varepsilon N^2 \int \limits_0^{2 \pi}d\phi \int \limits_{-L/2}^{L/2}dz \int \limits_0^{R}d\rho\ \rho J_m^2(\kappa \rho ) = \cr && \displaystyle = 2\varepsilon 2\pi L N^2 \int \limits_0^{R}d\rho\ \rho J_m^2(\kappa \rho ).   \label{Eq0000-6}
\end{eqnarray}
The integral over $\rho$ is evaluated with the use of Lommel formula \cite{Tr} (it can also be derived from Eq. 6.521.1 of \cite{GR}):
\begin{eqnarray}
& \displaystyle \int \limits_0^{R}d\rho\ \rho J_m^2(\kappa \rho ) = \cr & \displaystyle = \frac{R^2}{2} \Big ( (J_m^{\prime}(\kappa R))^2 + J_m^2 (\kappa R) \Big (1 - \frac{m^2}{\kappa^2 R^2}\Big )\Big ), \label{Eq0000-7}
\end{eqnarray}
As a result, we find the following expression for the normalization constant
\begin{eqnarray}
\displaystyle N = \frac{1}{\sqrt{2 \varepsilon V}} \frac{1}{\sqrt{  (J_m^{\prime}(\kappa R))^2 + J_m^2 (\kappa R) \Big (1 - \frac{m^2}{\kappa^2 R^2}\Big )}}, \label{Eq0000-8}
\end{eqnarray}

\begin{figure*}
\center
\includegraphics[width=18.00cm, height=4.00cm]{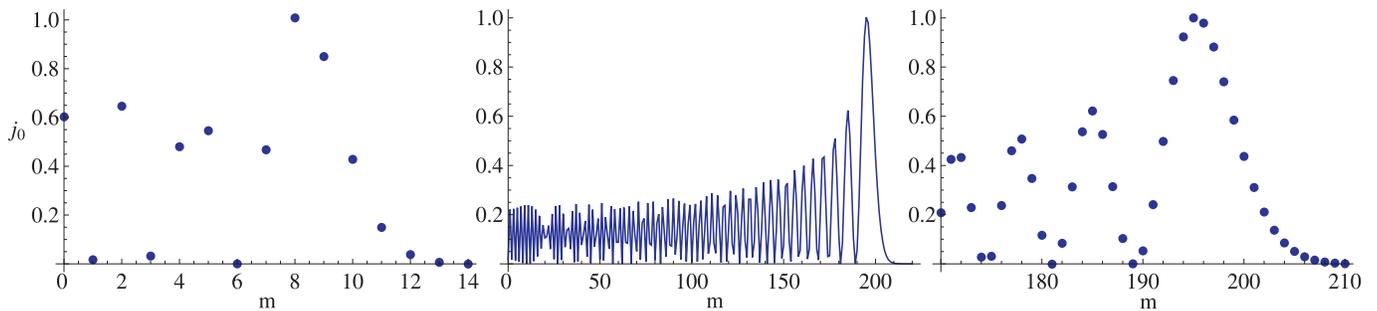}
\caption{\label{Fig0} The probability density (in arbitrary units) of a scalar for a given $\rho \kappa$ as a function of the OAM value. The left panel: $\rho \kappa = 10$; the central and right panels: $\rho \kappa = 200$. The points are joined with the straight lines in order to illustrate the quasi-continuous limiting case $\rho \kappa \gtrsim m_c  \gg \Delta m = 1$. Everywhere $R = 100 \rho$.}
\end{figure*}

For a given value of $\rho \kappa$, the probability density $j^0$ of this scalar state as a function of $m$ gets the maximum at $m_c \lesssim \rho \kappa$. In the limiting case when $\rho \kappa$ is greater than the $\Delta m = 1$ interval, the distribution over $m$ becomes almost continuous, see Fig.\ref{Fig0}. Mathematically, such a behavior is somewhat similar to the spectral distribution of the synchrotron radiation (just due to the same $J_m^2$ dependence; see e.g. \cite{Sokolov}).

For the large values of the cylinder's radius ($R \gg \kappa^{-1}$), we have (compare with \cite{Serbo, Ivanov}):
\begin{eqnarray}
\displaystyle \int \limits_0^{R\rightarrow \infty}d\rho\ \rho J_m^2(\kappa \rho ) \rightarrow \frac{R}{\pi \kappa},\ N \rightarrow \sqrt{\frac{\kappa}{4 \varepsilon L R}}. \label{Eq0000-9}
\end{eqnarray}
We now return to the problem of transition from the twisted states $|\psi \rangle$ to the plane-wave states in the limiting case $\kappa \rightarrow 0$. As can be seen from the last expression, a correct transition can be performed when getting rid of $\sqrt{\kappa}$ from the numerator of $N$:
\begin{eqnarray}
& \displaystyle |p\rangle : = \sqrt{\frac{2}{\pi R}} \frac{1}{\sqrt{\kappa}} |\varepsilon, p_{\parallel }, \kappa\rightarrow 0, m\rangle = \cr & \displaystyle = \frac{1}{\sqrt{2\varepsilon \pi R^2 L}} e^{-i\varepsilon t + i p_{\parallel }z}. \label{Eq0000-10}
\end{eqnarray}
It is this approach, which was used in the papers \cite{Serbo, Ivanov, I-S}. Such a fact has appeared because the formula (\ref{Eq0000-9}) itself is correct for non-zero transverse momenta $\kappa \gg R^{-1}$ only. When normalizing the twisted states in a finite volume (i.e. for the constant $N$ from (\ref{Eq0000-8})), such an additional multiplication is superfluous. Indeed, expanding the Bessel function for the small values of momenta, $\kappa \ll R^{-1}$, we have
\begin{eqnarray}
\displaystyle \sqrt{  (J_m^{\prime}(\kappa R))^2 + J_m^2 (\kappa R) \Big (1 - \frac{m^2}{\kappa^2 R^2}\Big )} \approx \delta_{m, 0} + O(\kappa). \label{Eq0000-11}
\end{eqnarray}
Taking into account Eq.(\ref{Eq0000}), we finally arrive at the ordinary plane-wave state
\begin{widetext}
\begin{eqnarray}
& \displaystyle |\varepsilon, p_{\parallel }, \kappa, m\rangle = \frac{1}{\sqrt{2 \omega V}} \frac{J_m (\rho \kappa ) e^{-i \varepsilon t + i p_{\parallel } z + i m\phi_r}}{\sqrt{  (J_m^{\prime}(\kappa R))^2 + J_m^2 (\kappa R) \Big (1 - \frac{m^2}{\kappa^2 R^2}\Big )}} \rightarrow \frac{1}{\sqrt{2\varepsilon V}} e^{-i\varepsilon t + i p_{\parallel }z}. \label{Eq0000-12}
\end{eqnarray}
\end{widetext}
with the normalization constant being equal to $N = 1/\sqrt{2\varepsilon V}$.

We would like to stress that when considering the twisted photons, a somewhat ``classical'' interpretation of the transverse momentum spread (non-zero $\kappa$s) is often given. Namely, they speak of rotation of the Poynting vector around the propagation axis (corkscrew-like energy trajectory). Whereas this may be useful for photon beams (see e.g. \cite{Berry}), for a single particle (especially, for a massive electron) such an interpretation is somewhat misleading, because these rotations of the momentum vector in a free space should result in the radiation of electromagnetic waves (similar to the undulator radiation). 

Actually no radiation is emitted by a free twisted electron, since when choosing the Fourier transform according to (\ref{Eq00}) the azimuthal angle of the momentum, $\phi_p$, stays undetermined and we integrate over it. This implies some quantum broadening of the classical rectilinear trajectory in the transverse plane rather than rotations in the classical sense. We shall return to this discussion later.

\section{\label{Sect2} Free twisted electron} 

A free electron with a projection of OAM onto the propagation axis is described via solution of the Dirac equation $(i\gamma \partial - m_e) \psi (r) = 0$ that can be sought in the way analogous to (\ref{Eq0}):
\begin{eqnarray}
\displaystyle \psi (r) = \int \frac{d^4 p}{(2 \pi)^4}\ \psi (p) N u (p) e^{-ipr}, \label{Eqa0}
\end{eqnarray}
In the standard representation of Dirac matrices $\gamma^{\mu}$, the bispinor $u(p)$, which obeys the Lorentz-invariant normalization condition $\bar{u} u = 2 m_e$ (here: $\bar{u} =  u^{\dagger} \gamma^0$), can be written as follows (see, for example, \cite{BLP}):
\begin{eqnarray}
\displaystyle u (p) = (\sqrt{\varepsilon  + m_e}\ w, \sqrt{\varepsilon - m_e} ({\bm n} {\bm \sigma}) w )^T. \label{Eqa1}
\end{eqnarray}
Here, ${\bm \sigma}$ are the Pauli matrices. The two-component spinor $w$, which obeys the normalization condition $w^\dagger w = 1$, can determine the helicity states of a free electron by the following equation: 
\begin{eqnarray}
& \displaystyle \frac{1}{2}({\bm n}{\bm \sigma}) w^{(\lambda )} = \lambda w^{(\lambda )},\ \lambda \pm 1/2,\cr & \displaystyle {\bm n} = {\bm p}/p = \{\sin \theta_p \cos \phi_p, \sin \theta_p \sin \phi_p, \cos \theta_p\}.\nonumber 
\end{eqnarray}
The solution for this equation can be written as follows:
\begin{eqnarray}
& \displaystyle w^{(\lambda )} ({\bm n}) = \frac{1}{\sqrt{2}}\Big (2 \lambda \sqrt{1 + 2 \lambda \cos \theta_p} e^{-i \phi_p/2}, \cr & \displaystyle \sqrt{1 - 2 \lambda \cos \theta_p} e^{i \phi_p/2}\Big )^T e^{i \lambda \phi_p}, \label{Eqa3}
\end{eqnarray}
where the common phase factor is chosen so that the azimuthal dependence vanishes in the paraxial limit $\theta_p \rightarrow 0$: 
\begin{eqnarray}
& \displaystyle
w^{(1/2)} (\theta_p \rightarrow 0) = (1, 0)^T,\ w^{(-1/2)} (\theta_p \rightarrow 0) = (0, 1)^T. \nonumber 
\end{eqnarray}
Note that the bispinor $u(p)$ can be represented in the following way
\begin{eqnarray}
\displaystyle u(p) \equiv u^{(1/2)} e^{i\phi_p (\lambda - 1/2)} + u^{(-1/2)} e^{i\phi_p (\lambda + 1/2)}. \label{Eq4a}
\end{eqnarray}

Choosing the function $\psi (p)$ in the form analogous to (\ref{Eq00}), but with the factors $(-i)^{m + \lambda \pm 1/2}$ instead of just $(-i)^m$, we get the Bessel states of the free electron, which were obtained in a bit different form in \cite{Leary, Bliokh-11}:
\begin{eqnarray}
& \displaystyle \psi (r) = N e^{-i \varepsilon t + i p_{\parallel} z } \Big (u^{(1/2)} J_{m + \lambda - 1/2} (\rho \kappa ) e^{i \phi_r (m + \lambda - 1/2)} \cr & \displaystyle + u^{(-1/2)} J_{m + \lambda + 1/2} (\rho \kappa ) e^{i \phi_r (m + \lambda + 1/2)} \Big ). \label{Eqa5}
\end{eqnarray}
The bispinors
\begin{widetext}
\begin{eqnarray}
& \displaystyle u^{(-1/2)} = \frac{1}{\sqrt{2}}\Big (0, \sqrt{\varepsilon + m_e}\sqrt{1 - 2 \lambda \cos \theta}, 0, 2\lambda \sqrt{\varepsilon - m_e} \sqrt{1 - 2 \lambda \cos \theta}\Big )^T,\cr & \displaystyle u^{(1/2)} = \frac{1}{\sqrt{2}}\Big (2 \lambda \sqrt{\varepsilon + m_e} \sqrt{1 + 2 \lambda \cos \theta}, 0, \sqrt{\varepsilon - m_e} \sqrt{1 + 2 \lambda \cos \theta}, 0\Big )^T \label{Eqa6}
\end{eqnarray}
\end{widetext}
are the eigenfunctions of the spin's $z$-projection operator with the eigenvalues $s_z = \pm 1/2$:
\begin{eqnarray}
& \displaystyle \hat{s}_z u^{(\pm 1/2)}: = \frac{1}{2} \Sigma_3 u^{(\pm 1/2)} = \pm \frac{1}{2} u^{(\pm 1/2)}, \label{Eqa7}
\end{eqnarray}
where in the standard representation ${\bm \Sigma} = \diag ({\bm \sigma}, {\bm \sigma})$. Here, $\theta$ is the polar cone angle: $\sin \theta = \kappa/\sqrt{p_{\parallel }^2 + \kappa^2}$.

The twisted state (\ref{Eqa5}) can be re-written as follows
\begin{widetext}
\begin{eqnarray}
& \displaystyle | \varepsilon, p_{\parallel }, \kappa, j_z = l_z + s_z = m + \Lambda \rangle  = |...l_z = m + \Lambda - 1/2, s_z = +1/2\rangle + |...l_z = m + \Lambda + 1/2, s_z = -1/2\rangle , \label {Eqa8} 
\end{eqnarray}
\end{widetext}
that means the Bessel state (\ref{Eqa5}) is an eigenfunction of $\hat{j}_z = \hat{l}_z + \hat{s}_z$ operator with an eigenvalue $m + \lambda$. Unlike the wave function of a scalar particle, this state represents a sum of two components that illustrates the spin-orbital connection of the free electron. Note that the operators $\hat{l}_z, \hat{s}_z$ do not commute separately with the Dirac Hamiltonian, but the sum of them $\hat{j}_z = \hat{l}_z + \hat{s}_z$ commutes that means the z-projection of the total angular momentum is an integral of motion. In the paraxial case, the operator $\hat{j}_z$ transforms into the one of helicity.

In order to calculate the density of the probability current for twisted states, one needs to calculate the products of the form $\bar{u}_{\pm 1/2}\gamma^{\mu}u_{\pm 1/2}$. By the direct calculation in the standard representation of Dirac matrices one can obtain the following expressions:
\begin{widetext}
\begin{eqnarray}
&& \displaystyle \bar{u}_{1/2}\gamma^{\mu}u_{1/2} = \{\varepsilon, 0, 0, 2 \lambda p\} (1 + 2 \lambda \cos \theta) \equiv p_{(\lambda)} (1 + 2 \lambda \cos \theta),\cr && \displaystyle \bar{u}_{1/2}\gamma^{\mu}u_{-1/2} = \kappa \{ 0, 1, -i, 0\} \equiv \kappa e_{(-)},\ \kappa = p \sin \theta, \cr && \displaystyle 
\bar{u}_{-1/2}\gamma^{\mu}u_{1/2} = \kappa \{ 0, 1, i, 0\} \equiv \kappa e_{(+)},\cr && \displaystyle
\bar{u}_{-1/2}\gamma^{\mu}u_{-1/2} = \{\varepsilon, 0, 0, - 2 \lambda p\} (1 - 2 \lambda \cos \theta) \equiv p_{(-\lambda)} (1 - 2 \lambda \cos \theta).
\label{Eqb12}
\end{eqnarray}
\end{widetext}
It is easy to check that
\begin{eqnarray}
& \displaystyle \bar{u}\gamma^{\mu}u  \equiv \bar{u}_{1/2}\gamma^{\mu}u_{1/2} + \bar{u}_{1/2}\gamma^{\mu}u_{-1/2} e^{i\phi_p} + \cr & \displaystyle + \bar{u}_{-1/2}\gamma^{\mu}u_{1/2}e^{-i\phi_p} + \bar{u}_{-1/2}\gamma^{\mu}u_{-1/2} = \cr & \displaystyle  = 2 p^{\mu} \equiv 2 \{\varepsilon, \kappa \cos \phi_p, \kappa \sin \phi_p, p \cos \theta\},
\label{Eqb13}
\end{eqnarray}
as should be.

Thus, for the current density one obtains the following formula (arguments of the Bessel functions are omitted):
\begin{widetext}
\begin{eqnarray}
& \displaystyle j^{\mu}({\bm \rho}) \equiv \bar{\psi}\gamma^{\mu}\psi = N^2 \Big ( J_{m + \lambda -1/2}^2 p_{(\lambda )}^{\mu} (1 + 2\lambda \cos \theta ) + 2 J_{m + \lambda -1/2}J_{m + \lambda +1/2} \kappa^{\mu}+ J_{m + \lambda +1/2}^2 p_{(-\lambda )}^{\mu} (1 - 2\lambda \cos \theta ) \Big ),
\label{Eqb14}
\end{eqnarray}
\end{widetext}
where 
\begin{eqnarray}
& \displaystyle \kappa^{\mu} = \{0, \kappa \cos \phi_r, \kappa \sin \phi_r, 0\} \nonumber
\end{eqnarray}
is denoted. 

This current density acquires the transverse components as well as a dependence upon the transverse coordinates. Due to the properties of the Bessel functions, the radial distribution of the probability density over $\rho$ has an oscillating character with the central minimum resembling the doughnut spatial structure of the focused photon beams (see e.g. figures in the second article of \cite{Serbo}). Such a feature for the particle is a pure quantum effect and it is sometimes called the OAM-Zitterbewegung \cite{Bliokh-07}. Nevertheless, in our view it is more correctly to speak of the OAM-induced quantum broadening of the classical trajectory than of the effective spiral trajectories of a free electron, because the last would mean the necessity of some undulator-like radiation, which is forbidden in actual fact. In addition, the current density (\ref{Eqb14}) does not depend on time similar to a plane-wave free electron that does not allow one to interpret the current's transverse components as some vibrations in the classical sense.

As can be checked, the set of functions (\ref{Eqa5}) is also orthogonal and complete:
\begin{eqnarray}
& \displaystyle \int d^3 r\ \psi^{*}_{\varepsilon, p_{\parallel }^{\prime}, \kappa^{\prime}, m^{\prime}, \lambda^{\prime}} (r) \psi_{\varepsilon, p_{\parallel }, \kappa, m, \lambda} (r) = \cr & \displaystyle = N^2 (2 \pi )^2 2 \varepsilon \delta (p_{\parallel} - p_{\parallel}^{\prime}) \frac{\delta (\kappa - \kappa^{\prime})}{\kappa}\delta_{\lambda \lambda^{\prime}} \delta_{m m^{\prime}}, \cr & \displaystyle \sum \limits_{m} \int \frac{d^3 p}{(2 \pi )^3}\ \psi^{*}_{\varepsilon, p_{\parallel }, \kappa, m, \lambda} (r^{\prime}) \psi_{\varepsilon, p_{\parallel }, \kappa, m, \lambda} (r) = \cr & \displaystyle = N^2 2 \varepsilon \delta ({\bm r} - {\bm r}^{\prime}). \label{Eq10b}
\end{eqnarray}

We determine the normalization constant in (\ref{Eqa5}) with the condition $1 = \int d^3 r\ j^0,\ j^0 = |\psi(r)|^2$ that leads to the following expression
\begin{eqnarray}
& \displaystyle N = \frac{1}{\sqrt{\varepsilon V}} \Big ( (1 + 2 \lambda \cos \theta ) \mathcal {J}_{m + \lambda - 1/2} + \cr & \displaystyle + (1 - 2 \lambda \cos \theta ) \mathcal {J}_{m + \lambda + 1/2}\Big )^{-1/2}, \label{Eqa11}
\end{eqnarray}
where $\mathcal {J}_{m + \lambda \pm 1/2}$ is defined as follows (see Eq.(\ref{Eq0000-7}))
\begin{eqnarray}
& \displaystyle & \displaystyle \mathcal {J}_m (\kappa R) : = (J_m^{\prime}(\kappa R))^2 + J_m^2 (\kappa R) \Big (1 - \frac{m^2}{\kappa^2 R^2}\Big ) \label{Eq1000000000}
\end{eqnarray}
Note also that in the paraxial limit ($\theta \rightarrow 0$) the twisted state transforms into the plane wave with $m = 0$, and the normalization constant in this case is $N \rightarrow 1/\sqrt{2\varepsilon V}$.

\section{\label{Sect3} Twisted electron in a plane electromagnetic wave} 

The twisted states of fermions in external potentials were previously studied in \cite{O, Leary, Bagrov}. For a field of a plane electromagnetic wave with a potential $A^{\mu} = A^{\mu} (\varphi), \varphi = k^{\mu}r_{\mu}$, the so-called Volkov states are usually used (see e.g. \cite{BLP, R}), which can be called the plane-wave states, since they transform to the free electron's plane-wave ones when the external field is switched off. Here we construct the twisted states of an electron moving in a plane electromagnetic wave. The existence of such ``non-Volkov solutions'' for the Klein-Gordon and Dirac equations was indicated in \cite{Bagrov}. 

Let the wave move in the negative z-direction, 
\begin{eqnarray}
& \displaystyle k = \{\omega, 0, 0, - \omega\} \equiv \omega n,\ n = \{1, 0, 0, -1\},\cr & \displaystyle \varphi = kr = \omega (t + z), \nonumber
\end{eqnarray} 
and it is adiabatically switched off in the far past and future. Let the twisted electron move in the positive z-direction on average, whereas it does strictly along the z-axis in the paraxial case when $\kappa \rightarrow 0$. We shall choose the potential of the laser wave in the usual way (see e.g. \cite{BLP, R}). Namely, we imply the Lorentz gauge that in the case under consideration means $k A = \omega (A^0 + A^3) = 0$, and we also suppose that potential is a space-like four-vector: $A^2 < 0$. These suggestions completely fix the gauge.

The wave functions of electron with some z-projection of OAM can be sought as the integral over the plane-wave Volkov solutions of the Dirac equation:
\begin{eqnarray}
& \displaystyle |\psi \rangle = \int \frac{d^4 p}{(2 \pi)^4}\ \psi (p) |p\rangle, \cr & \displaystyle |p\rangle = \psi_V (r) = N \Big ( 1 + \frac{e}{2(pk)} (\gamma k) (\gamma A)\Big ) u(p) e^{i S},  \cr & \displaystyle S = -(pr) - \frac{e}{(pk)} \int d \varphi\ \Big ((pA) - \frac{e}{2} A^2 \Big ).
\label{Eq4.1}
\end{eqnarray}
Here, $S$ is the classical action for the electron in the wave, while the bispinor
\begin{eqnarray}
\displaystyle \Big ( 1 + \frac{e}{2(pk)} (\gamma k) (\gamma A)\Big ) = \exp \Big \{\frac{e}{4(pk)} \int d\varphi\ F_{\mu \nu} \sigma^{\mu \nu}\Big \}
\label{Eq4.2}
\end{eqnarray}
takes into account an interaction of the electron's spin with the wave's field. Here, $F^{\mu \nu} = \partial^{\mu} A^{\nu} - \partial^{\nu} A^{\mu}$ and $\sigma^{\mu \nu} = (\gamma^{\mu} \gamma^{\nu} - \gamma^{\nu} \gamma^{\mu})/2$ are denoted. The constant vector $p$, which has all four components, in the limiting case $\kappa \rightarrow 0, A\rightarrow 0$ transforms to the 4-momentum of a free plane-wave electron $\{\varepsilon, 0, 0, p\}$.

Choosing the function $\psi (p)$ is the form analogous to (\ref{Eq00}) (remember the factors $(-i)^{\pm 1/2}$, which are hidden in $u(p)$), we obtain the following intermediate result:
\begin{widetext}
\begin{eqnarray}
& \displaystyle \psi (r) = N (-i)^{m + \lambda} \exp\Big \{ -i \varepsilon t + ip_{\parallel }z + i \frac{e^2}{2(pk)} \int d\varphi\ A^2\Big \} \Big ( 1 + \frac{e}{2(pk)} (\gamma k) (\gamma A)\Big ) \cr & \displaystyle \times \int \limits_0^{2\pi} \frac{d \phi_p}{2\pi}\ u(p) \exp \Big \{im\phi_p + i\kappa \rho \cos (\phi_r - \phi_p ) - i\frac{e}{(pk)}\int d \varphi\ (pA)\Big \}.
\label{Eq4.3}
\end{eqnarray}
\end{widetext}
Since $p^{\mu}$ has the transverse components, the product $(pA)$ may depend upon the integration variable $\phi_p$. Indeed, all the terms depending upon the azimuthal angle in the action (\ref{Eq4.1}) are:
\begin{eqnarray}
& \displaystyle -(pr) - \frac{e}{(pk)} \int d \varphi\ (pA) \equiv - (p \mathcal R),\cr & \displaystyle \mathcal R^{\mu} = r^{\mu} + \frac{e}{(pk)} \int d \varphi\ A^{\mu}.
\label{Eq4.4}
\end{eqnarray}
Note that if the time averaging is applied, one has:
\begin{eqnarray}
& \displaystyle \bar {\mathcal R} = r.
\nonumber
\end{eqnarray}
It is clear already from this ratio that the electron in the wave may not have a definite z-projection of the OAM (or even of the total momentum: OAM + spin) and it may have only the time-averaged ``effective'' angular momentum. This situation is analogous to the existance of the mean quasi-momentum of the electron without OAM in the wave's field (see, for example, \cite{BLP, R} and below).

In order to evaluate the integrals in (\ref{Eq4.3}), (\ref{Eq4.4}), it is necessary to define the model of the plane wave. Let us consider for definiteness a 100 $\%$ circularly polarized laser wave with the following potential
\begin{eqnarray}
& \displaystyle A = a \{0, \cos \varphi, \sin \varphi ,0\},\ A^2 = - a^2.
\label{Eq4.6}
\end{eqnarray}
Thus, we have:
\begin{eqnarray}
& \displaystyle im\phi_p + i\kappa \rho \cos (\phi_r - \phi_p ) - i\frac{e}{(pk)}\int d \varphi\ (pA) = \cr & \displaystyle = im\phi_p + i \kappa \mathcal R_{\perp } \cos (\phi_{\mathcal R} - \phi_p), \cr & \displaystyle \mathcal R_{\perp }^2 = \Big (r_{\perp } + \frac{e}{(pk)} \int d \varphi\ A_{\perp }\Big )^2, \cr & \displaystyle \mathcal R_{\perp }^{\mu} = \{0, \mathcal R_{\perp } \cos \phi_{\mathcal R}, \mathcal R_{\perp } \sin \phi_{\mathcal R}, 0\}.
\label{Eq4.7}
\end{eqnarray}
Inserting this into Eq.(\ref{Eq4.3}), one finally obtains the following wave function:
\begin{widetext}
\begin{eqnarray}
& \displaystyle \psi (r) = N \exp\Big \{ -i \varepsilon t + ip_{\parallel }z - i \frac{e^2 a^2}{2(pk)}(kr) \Big \} \Big ( 1 + \frac{e}{2(pk)} (\gamma k) (\gamma A)\Big ) \cr & \displaystyle \times \Big (u^{(1/2)} J_{m + \lambda - 1/2} (\kappa \mathcal R_{\perp }) e^{i\phi_{\mathcal R} (m + \lambda -1/2)} + u^{(-1/2)} J_{m + \lambda +1/2} (\kappa \mathcal R_{\perp }) e^{i\phi_{\mathcal R} (m + \lambda + 1/2)}\Big ),
\label{Eq4.8}
\end{eqnarray}
\end{widetext}
which is in a close analogy with Eq.(\ref{Eqa5}). 

As can be seen, when the field of the wave vanishes, $A \rightarrow 0$, we are left with the Bessel states of the free electron (\ref{Eqa5}), whereas we have the ordinary Volkov solutions in the paraxial case ($\kappa \rightarrow 0$) \footnote{The term with $(pA)$ in the exponential vanishes because in this case $p \rightarrow \{\varepsilon, 0, 0, p\}$ and $A = \{0, A_x, A_y, 0\}$.}. Thus, the twisted states obtained differ from the free-electron twisted states (\ref{Eqa5}) in the replacement of the energy and the longitudinal momentum by the quasi-energy and the corresponding quasi-momentum (see their definitions below) and also in the replacement of the transverse coordinates: ${\bm \rho}\rightarrow {\bm {\mathcal R}}_{\perp }$. 

Note that the orthogonality and completeness of the states (\ref{Eq4.8}) follow already from the adiabatic switching-on and -off of the laser wave at $t \rightarrow \pm \infty$.

Before normalizing the states (\ref{Eq4.8}), let us calculate their current density. Calculation is performed with the use of Eq.(\ref{Eqb12}) and the identity
\begin{eqnarray}
& \displaystyle \Big ( 1 + \frac{e}{2(pk)} (\gamma A) (\gamma k)\Big ) \gamma^{\mu} \Big ( 1 + \frac{e}{2(pk)} (\gamma k) (\gamma A)\Big ) = \cr & \displaystyle = \gamma^{\mu} - eA^{\mu} \frac{(\gamma k)}{(pk)} + k^{\mu} \frac{e}{(pk)}\Big ((\gamma A) - \frac{e A^2 (\gamma k)}{2 (pk)}\Big ).
\label{Eq4.9}
\end{eqnarray}
As a result, we obtain the following expression (arguments of the Bessel functions are not shown):
\begin{widetext}
\begin{eqnarray}
& \displaystyle j^{\mu}(r) = \bar{\psi}\gamma^{\mu} \psi = N^2  \Bigg ( J_{m + \lambda -1/2}^2 (1 + 2\lambda \cos \theta ) \Big (p_{(\lambda )}^{\mu} - eA^{\mu} \frac{(k p_{(\lambda )})}{(pk)} - k^{\mu} \frac{e^2 A^2 (k p_{(\lambda )})}{2 (pk)^2}\Big ) +  \cr & \displaystyle 2 J_{m + \lambda -1/2}J_{m + \lambda +1/2} \Big (\kappa_{\mathcal R}^{\mu} + e k^{\mu} \frac{(\kappa_{\mathcal R} A)}{(pk)}\Big ) + J_{m + \lambda +1/2}^2 (1 - 2\lambda \cos \theta ) \Big (p_{(-\lambda )}^{\mu} - eA^{\mu} \frac{(k p_{(-\lambda )})}{(pk)} - k^{\mu} \frac{e^2 A^2 (k p_{(-\lambda )})}{2 (pk)^2}\Big )  \Bigg ) \label{Eq4.10}
\end{eqnarray}
\end{widetext}
where 
\begin{eqnarray}
& \displaystyle \kappa_{\mathcal R}^{\mu} = \{0, \kappa \cos \phi_{\mathcal R}, \kappa \sin \phi_{\mathcal R}, 0\}\nonumber
\end{eqnarray} 
is denoted. 

This current density reveals a twofold character of the electron's motion: the terms with $A^{\mu} (\varphi)$, which vanish on average, describe the helical quasi-classical motion of the wave-packet's center, whereas the non-paraxial terms with $\kappa_{\mathcal R}^{\mu}$ correspond to the purely quantum broadening of the classical trajectory due to OAM. In contrast to the free twisted electron, this current does depend on time, but this dependence has appeared solely due to the laser wave $A$. Thus, the electron's effective trajectory in this case is a classical spiral with some OAM-induced transverse momentum spread. Note that such an effect is somewhat similar to the well-known quantum broadening of the classical trajectory in synchrotron radiation due to emission of hard photons \cite{Sokolov}, but the difference is that in this case this spreading has no relation to the radiation process, since it remains even for a free twisted particle.

The mean value of (\ref{Eq4.9}) gives the electron's kinetic momentum in the wave $\pi^{\mu}$. Its time average is called the quasi-momentum $q^{\mu}:= \bar{\pi}^{\mu}$ (see \cite{BLP, R} in more detail):
\begin{eqnarray}
& \displaystyle \bar {u}\Big ( \gamma^{\mu} - eA^{\mu} \frac{(\gamma k)}{(pk)} + k^{\mu} \frac{e}{(pk)}\Big ((\gamma A) - \frac{e A^2 (\gamma k)}{2 (pk)}\Big )\Big ) u = \cr & \displaystyle = 2 \Big (p^{\mu} - e A^{\mu} + k^{\mu} \frac{e}{(pk)}\Big ((pA) - \frac{e A^2}{2}\Big )\Big ) \equiv 2\pi ^{\mu} \rightarrow \cr & \displaystyle  \rightarrow 2 \Big (p^{\mu} - k^{\mu} \frac{e^2 \bar{A^2}}{2 (pk)}\Big ) \equiv 2 q^{\mu}.
\label{Eq4.11}
\end{eqnarray}
It is clear from this expression that the terms at the Bessel functions in (\ref{Eq4.10}) represent the components of the electron's kinetic momentum. Here $\bar{A^2}$ denotes the time average of $A^2$. Note also that in terms of the vector $q^{\mu}$, the common exponent in (\ref{Eq4.8}) can be represented simply as $\exp \{ -i q_0 t + iq_{\parallel }z \}$.

We would like to stress the noticeable difference between the expressions obtained and the ones for a free twisted electron. Here, the arguments of the Bessel functions depend upon the plane wave's phase and, consequently, on time: $\mathcal R_{\perp } \equiv \mathcal R_{\perp } (\varphi)$. On the face of it, if we normalized the twisted states in a way analogous to the Volkov electron as $\int d^3 r\ \bar{j^0} = 1$ (see, for example, $\S$ 101 in \cite{BLP}), this fact would forbid the time averaging simply by striking off all the terms linear on the wave's phase $\varphi$. Nevertheless, thanks to the fact that $\varphi = (kr) = (k\mathcal R) = \omega (\mathcal R^0 + \mathcal R^3)$, the Bessel functions standing under the integral over coordinates can be considered to be independent of $\mathcal R^0$ that allows one to perform the time averaging in the usual way. Moreover, for the model of the laser wave (\ref{Eq4.6}) the time averaging is actually unnecessary, since $A^2 = -a^2$ does not depend on time and $A^0 = 0$. We shall return to this problem a bit later. 

Finally, it is easy to check that in the case under consideration: $d^3 r = d^3 \mathcal R$ (the Jacobian equals unity), so for the normalization constant we get:
\begin{widetext}
\begin{eqnarray}
& \displaystyle \int \limits_V d^3 r\ j^0 = \int \limits_V d^3 \mathcal R\ j^0 = N^2 2 \pi L \int \limits_0^{R} d \mathcal R_{\perp }\ \mathcal R_{\perp } \Big (J_{m + \lambda - 1/2}^2 (1 + 2\lambda \cos \theta ) q^0_{(\lambda)} + J_{m + \lambda + 1/2}^2 (1 - 2\lambda \cos \theta ) q^0_{(-\lambda)} \Big ) = 1 \cr & \displaystyle  \Rightarrow N = \frac{1}{\sqrt{V}}\Big (\mathcal {J}_{m + \lambda - 1/2} (1 + 2\lambda \cos \theta ) q^0_{(\lambda)} + \mathcal {J}_{m + \lambda + 1/2} (1 - 2\lambda \cos \theta ) q^0_{(-\lambda)} \Big )^{-1/2}.
\label{Eq4.12}
\end{eqnarray}
\end{widetext}
Here, $\mathcal {J}_m$ (which is not a Bessel function) is defined in (\ref{Eq1000000000}) and we have also introduced the denotations:
\begin{eqnarray}
& \displaystyle q^0_{(\lambda)} = \varepsilon - \omega \frac{e^2 A^2 (kp_{(\lambda )})}{2(pk)^2}, \cr & \displaystyle q^0_{(-\lambda)} = \varepsilon - \omega \frac{e^2 A^2 (kp_{(-\lambda )})}{2(pk)^2}.
\label{Eq4.13}
\end{eqnarray}
In the paraxial limit ($\theta \rightarrow 0$), we have the ordinary Volkov state with $m = 0$ and the normalization constant is $N \rightarrow 1/\sqrt{2 q^0 V}$, as expected.

As in the case of a free twisted electron, the state (\ref{Eq4.8}) represents a sum of two functions. Each of these terms is an eigenfunction for the operator $- i \partial_{{\phi}_{\mathcal R}}$, but the differentiation now is done over the azimuthal angle of the vector $\mathcal {R}_{\perp }$ instead of the one of ${\bm \rho}$. As easy to show, it is the operator $\hat{L}_z = -i \partial_{\phi_{\mathcal R}}$, which is an integral of motion for a scalar particle in the laser field instead of $\hat{l}_z = - i \partial_{\phi_r}$ \cite{Bagrov}. Let us calculate for simplicity the mean value of these operators by using the corresponding solutions of the Klein-Gordon equation:
\begin{eqnarray}
& \displaystyle \psi (r) = N \exp \{ -i q_0 t + iq_{\parallel }z\} J_m (\kappa \mathcal R_{\perp }) e^{i\phi_{\mathcal R} m}.
\label{Eq4.8KG}
\end{eqnarray}
Note that this solution has a form of the one of a free twisted electron (compare with (\ref{Eq000}); see also \cite{Bagrov} and discussion on this in Sec.\ref{Sect6} hereinafter). After rewriting the operator $\hat{l}_z = - i \partial_{\phi_r}$ in terms of new variables
\begin{eqnarray}
& \displaystyle \hat{l}_z = - i \frac{\partial}{\partial \phi_r} = - i \Big ( 1 + \frac{e a}{(pk)} \frac{\sin (\phi_{\mathcal R} - \varphi )}{{\mathcal R}_{\perp }}\Big ) \frac{\partial}{\partial \phi_{\mathcal R}} + \cr & \displaystyle + i \frac{e a}{(pk)} \cos (\phi_{\mathcal R} - \varphi )\frac{\partial}{\partial {\mathcal R}_{\perp }},
\label{Eq4.14}
\end{eqnarray}
we arrive at:
\begin{eqnarray}
& \displaystyle \langle \hat{l}_z \rangle = \int d^3 \mathcal R\ \psi^{*}(r) \hat{l}_z \psi (r)\Big /\int d^3 \mathcal R\ |\psi (r)|^2 = \cr & \displaystyle = m = \langle \hat{L}_z \rangle.
\label{Eq4.15}
\end{eqnarray}
All the ``superfluous'' terms here have vanished after the integration over $\phi_{\mathcal R}$.

\section{\label{Sect4} Spin of a plane-wave electron in a strong laser wave} 

The twisted state obtained describes the electron with an ``almost'' definite OAM due to the spin-orbit coupling. Let us take a closer look at the spin characteristics of this state. It is easy to see that when the free-electron bispinor $u (p)$ is a helicity state with ${\bm p} = \{0, 0, p\}$, the Volkov state is not: 
\begin{eqnarray}
& \displaystyle \Sigma_z \psi_V(r) \ne 2\lambda \psi_V (r). \nonumber
\end{eqnarray}
An analogous situation takes place for the twisted state (\ref{Eq4.8}): action of the $\Sigma_z$ operator does not lead to the simple multiplication on $\pm 1$. To begin with, let us calculate the mean value of the spin's z-projection over the ordinary Volkov states without OAM:
\begin{widetext}
\begin{eqnarray}
& \displaystyle \langle \hat{s}_z \rangle = \int d^3 r\ \psi_V^* (r) \frac{1}{2} \Sigma_z \psi_V (r) = N^2 \int d^3 r\ \bar {u} \Big ( 1 + \frac{e}{2(pk)} (\gamma A) (\gamma k) \Big ) \gamma^0 \frac{1}{2}\Sigma_z \Big ( 1 + \frac{e}{2(pk)} (\gamma k) (\gamma A)\Big ) u
\label{Eq5.1}
\end{eqnarray}
\end{widetext}
Noting that for the wave model under consideration the spin exponential (\ref{Eq4.2}) contains only the matrices \footnote{$F_{\mu \nu} \sigma^{\mu \nu} = 2 ({\bm \alpha} {\bm E}) + 2i ({\bm \Sigma}{\bm H})$} $\Sigma_{x, y}, \alpha_{x, y}$, which all anti-commute with $\Sigma_z$, we obtain the simple expression:
\begin{eqnarray}
& \displaystyle \hat{s}_z \psi_V (r) = \lambda \exp\Big \{- \frac{e}{2(pk)} \int d\varphi\ F_{\mu \nu} \sigma^{\mu \nu}\Big \} \psi_V (r) = \cr & \displaystyle = \lambda \Big ( 1 + \frac{e}{(pk)} (\gamma A) (\gamma k) \Big ) \psi_V (r).
\label{Eq5.10}
\end{eqnarray}
Multiplying this on $\psi_V^*$ from the left, then taking into account the following easily verifiable formulas (remember that $ {\bm \alpha} = -\gamma^5 {\bm \Sigma}, \gamma^5 {\bm \Sigma} = {\bm \Sigma} \gamma^5$ \cite{Bagrov-02, BLP})
\begin{eqnarray}
& \displaystyle u^* {\bm \alpha} u = 2 {\bm p},\ u^* \Sigma_z u = 4 \varepsilon \lambda,\ u^*\gamma^5 u = - 4 p \lambda, \label{Eq5.7}
\end{eqnarray} 
and performing the time averaging, we arrive at the following expression for the mean value of the effective spin (remember that $N^2 = 1/2 q_0 V$):
\begin{eqnarray}
& \displaystyle \langle \bar {\hat{s}}_z \rangle = N^2 \int d^3 r\ \lambda \Big ( 2\varepsilon - \omega \frac{e^2 a^2}{(pk)} \Big ) = \lambda \frac{ 2\varepsilon - \omega \frac{e^2 a^2}{(pk)}}{ 2\varepsilon + \omega \frac{e^2 a^2}{(pk)}}.
\label{Eq5.11}
\end{eqnarray}
This formula means that the electron's spin in a laser wave experiences precession with some mean value of the helicity, which is shifted due to the laser wave (in analogy with the quasi-momentum). It is interesting that such a shift occurs to the ``wrong'' direction resulting in reduction of the effective spin. In the formal limiting case of the strong laser wave with $a^2 \rightarrow \infty$, we have a spin inversion (non-radiative spin-flip): $\langle \bar {\hat{s}}_z \rangle \rightarrow -\lambda$.

A difference of $ \langle \bar {\hat{s}}_z \rangle$ from $\lambda$ may be caused by two reasons: the change of the electron polarization (the effective spin invariant $s^2$ stays to be equal to $-1$), or by the effective depolarization of the electron ($|s^2| < 1$). Therefore, let us study a similar shift for the invariant $s^2$ of the spin four-vector (by analogy with the mass shift). The general expression for the effective polarization vector in a laser wave is calculated as a mean value of $\gamma^{\mu} \gamma^5$ over the Volkov states \cite{R}:
\begin{eqnarray}
& \displaystyle {s_*}^{\mu} = s^{\mu} - \frac{e}{(pk)} \Big (A^{\mu} (sk) - k^{\mu} (sA)\Big ) - \cr & \displaystyle - k^{\mu} \frac{e^2 A^2}{2 (pk)^2} (sk). \label{Eq5.12}
\end{eqnarray}

Indeed, without a time averaging the square of the electron's kinetic momentum $\pi^{\mu}$ (see (\ref{Eq4.11})) coincides with the square of the free-electron's momentum and, consequently, with its ``bare'' mass: $\pi^2 = p^2 = m_e^2$. However after the time averaging one gets an effective mass: 
\begin{eqnarray}
& \displaystyle \bar{\pi}^{\mu} = : q^{\mu},\ q^2 = m_*^2 : = m_e^2 \Big ( 1 - \frac{e^2 \bar{A^2}}{m_e^2} \Big ) > m_e^2. \nonumber 
\end{eqnarray}
An analogous situation takes place for the spin invariant: before the time averaging we have for helicity states just
\begin{eqnarray}
& \displaystyle {s_*}^2 = s^2 = - {\bm \zeta}^2 = - (2 \lambda)^2 = -1. \label{Eq5.14}
\end{eqnarray}
However after the averaging we arrive at the following
\begin{eqnarray}
& \displaystyle \bar{s_*}^{\mu} = s^{\mu} - k^{\mu} \frac{e^2 \bar{A^2}}{2 (pk)^2} (sk), \cr & \displaystyle {\bar{s_*}}^2 = s^2 - e^2  \bar{A^2} \frac{(sk)^2}{(pk)^2}. \label{Eq5.15}
\end{eqnarray}
For a head-on collision of a longitudinally polarized electron with a circularly polarized wave, we have (see components of the spin 4-vector e.g. in \cite{BLP}) 
\begin{eqnarray}
& \displaystyle \frac{(sk)^2}{(pk)^2} = \frac{(2\lambda)^2}{m_e^2} \equiv -\frac{s^2}{m_e^2}. \label{Eq5.15a}
\end{eqnarray} 
As a result, we arrive at the following spin invariant's shift, which is always negative (it is clear already from (\ref{Eq5.15})):
\begin{eqnarray}
& \displaystyle {\bar{s_*}}^2 = s^2 \Big (1 + \frac{e^2  \bar{A^2}}{m_e^2}\Big )\Rightarrow\cr & \displaystyle  \frac{1}{2} \sqrt {-{\bar{s_*}}^2} = \lambda \sqrt{1 - \eta^2},\ \eta^2 = - \frac{e^2 \bar{A^2}}{m_e^2}. \label{Eq5.16}
\end{eqnarray}
So in contrast to the mass shift, here 
\begin{eqnarray}
- {\bar{s_*}}^2 < - s^2 \nonumber
\end{eqnarray}
Actually, the shifts of the mass and spin occur in the different directions just because the momentum $p^{\mu}$ is a time-like four-vector, $p^2 > 0$, whereas the spin four-vector is space-like: $s^2 < 0$. 

As follows from these formulas, the electron's effective spin in a strong laser wave can formally turn into zero and, what is maybe even more unexpected, the invariant ${\bar{s_*}}^2$ changes its sign in the point $\eta = 1$. In fact, this means that in the classically strong laser fields with $\eta \gtrsim 1$ the electron (or the electron beam) can behave itself as being spontaneously depolarized. We would like to emphasize that such an effective reduction of spin is absolutely ``kinematic'' (because it appears due to precession) and it has no relation to the so-called radiative polarization (or the Sokolov-Ternov effect) arising due to the radiation of photons (see discussion on this process in a laser wave e.g. in \cite{I}). In other words, in the case under consideration the electron in the laser wave is absolutely ``stable'', so such a shift appears only when it moves in the wave.

On the other hand, for a realistic laser pulse of a \textit{finite duration} the electron's spin state after leaving the wave may not coincide with the one before entering the wave. In principal, this makes such a spin shift experimentally detectable (see also a recent discussion on the similar effects for a quantized laser wave in \cite{Sk}). Investigating the effective spin's z-component in the laboratory frame (see (\ref{Eq5.11})), we find that its mean value $\langle \bar {\hat{s}}_z \rangle$ turns into zero under the following condition:
\begin{eqnarray}
& \displaystyle \varepsilon = \omega \frac{e^2 a_0^2}{2 (pk)} = \eta_0^2 \frac{m_e^2}{2 (p + \varepsilon)}, \label{Eq5.17}
\end{eqnarray}
From this, we have ($\gamma = \varepsilon/m_e$):
\begin{eqnarray}
& \displaystyle \eta_0^2 = 2 \gamma^2 (1 + \beta ) \label{Eq5.18}
\end{eqnarray}
with the minimum value of $\eta_0^2$ equal to 2. Thus, for the lasers with $\eta \gtrsim 1$ such a shift may be distinguished for the non-relativistic electrons.

\section{\label{Sect5} Total angular momentum of the electron in a strong laser wave} 

Consider now the twisted electron in a laser wave. Let us calculate the mean value of the total angular momentum's operator 
\begin{eqnarray}
& \hat{j}_z = -i\partial_{\phi_{\mathcal R}} + \frac{1}{2}\Sigma_z \nonumber
\end{eqnarray} 
over the states (\ref{Eq4.8}). Taking into account Eqs.(\ref{Eqb12}) and (\ref{Eq4.9}), we arrive at the following expression for the mean value of OAM:
\begin{eqnarray}
& \displaystyle \langle \hat{L}_z \rangle = \int d^3 \mathcal R\ \psi^*(r) (-i\partial_{\phi_{\mathcal R}}) \psi (r) = \cr & \displaystyle = N^2 \int d^3 \mathcal R\ \Big ((m + \lambda - 1/2) q^0_{(\lambda)} (1 + 2 \lambda \cos \theta )\times \cr & \displaystyle \times J^2_{m + \lambda -1/2} + (m + \lambda + 1/2) q^0_{(-\lambda)} (1 - 2 \lambda \cos \theta ) \times \cr & \displaystyle \times J^2_{m + \lambda +1/2}\Big ) \label{Eq5.19}
\end{eqnarray}
which is in a close analogy with the free twisted electron. Here, we have omitted the terms vanishing after the integration over the azimuthal angle $\phi_{\mathcal R}$. When calculating the mean value of the spin's z-projection we use the following formula:
\begin{widetext}
\begin{eqnarray}
& \displaystyle\Big ( 1 + \frac{e}{2(pk)} (\gamma A) (\gamma k)\Big ) \gamma^0  \Big ( 1 + \frac{e}{2(pk)} (\gamma A) (\gamma k)\Big ) \equiv  \cr & \displaystyle \equiv \exp \Big \{- \frac{e}{4(pk)} \int d\varphi\ F_{\mu \nu} \sigma^{\mu \nu}\Big \} \gamma^0 \exp \Big \{- \frac{e}{4(pk)} \int d\varphi\ F_{\mu \nu} \sigma^{\mu \nu}\Big \} = \cr & \displaystyle = \gamma^0 \Big (1 - \frac{e^2 a^2 \omega^2}{2 (pk)^2} - \frac{ie}{(pk)} (\tilde{{\bm H}} {\bm \Sigma}) + \frac{ie^2}{2(pk)^2}  (\tilde{{\bm H}} {\bm \Sigma}) (\tilde{{\bm E}} {\bm \alpha})\Big ),\cr & \displaystyle \tilde{{\bm E}} := \int d\varphi\ {\bm E},\ \tilde{{\bm H}} := \int d\varphi\ {\bm H}. \label{Eq5.20}
\end{eqnarray}
\end{widetext}
When integrating the mean value of this expression over $\phi_{\mathcal R}$, the non-zero contribution comes solely from the terms independent of the azimuthal angle. Furthermore, the term $(\tilde{{\bm H}} {\bm \Sigma})$ does not make a contribution at all, since $u^*_{\pm 1/2} \Sigma_{x,y} u_{\pm 1/2} = 0$, as easy to check. Finally, the contribution of the last term in (\ref{Eq5.20}) is reduced to the calculation of the following averages:
\begin{eqnarray}
& \displaystyle u^*_{1/2}\gamma^5 u_{1/2} = - 2 \lambda p (1 + 2 \lambda \cos \theta),\cr & \displaystyle u^*_{-1/2}\gamma^5 u_{-1/2} = - 2 \lambda p (1 - 2 \lambda \cos \theta),\cr & \displaystyle u^*_{\pm 1/2}\gamma^5 u_{\mp 1/2} = 0. \label{Eq5.21}
\end{eqnarray}

As a result, we have (arguments of the Bessel functions are not shown):
\begin{eqnarray}
& \displaystyle \frac{1}{2} \langle\Sigma_z\rangle =  \frac{1}{2} N^2\int d^3 \mathcal R\ \Big ((1 + 2 \lambda \cos \theta ) J^2_{m + \lambda -1/2}\times \cr & \displaystyle \times \Big (\varepsilon - \omega \frac{e^2 a^2}{2 (pk)^2} (k p_{(\lambda )}) \Big ) - (1 - 2 \lambda \cos \theta ) \times \cr & \displaystyle \times J^2_{m + \lambda +1/2}\Big (\varepsilon - \omega \frac{e^2 a^2}{2 (pk)^2} (k p_{(-\lambda )}) \Big )\Big ). \label{Eq5.22}
\end{eqnarray}

If the terms $q^0_{(\pm \lambda)} = \varepsilon + \omega e^2 a^2 (k p_{(\pm \lambda )})/2 (pk)^2$ took place in this expression instead of $\varepsilon - \omega e^2 a^2 (k p_{(\pm \lambda )})/2 (pk)^2$ (see Eq.(\ref{Eq4.13})), we would obtain exactly $\langle \hat{j}_z \rangle = m + \lambda$ (as for the free twisted electron) when summing Eq.(\ref{Eq5.22}) up with $\langle \hat{L}_z \rangle$ from Eq.(\ref{Eq5.19}) (see the expression for the normalization constant in (\ref{Eq4.12})). However, as stated above, the laser wave results in the shift of the electron's spin in the direction opposite to the momentum. It leads to the following final expression for the z-projection of the total angular momentum:
\begin{widetext}
\begin{eqnarray}
& \displaystyle \langle \hat{j}_z \rangle = N^2 \int d^3 \mathcal R\ \Big ((1 + 2 \lambda \cos \theta ) J^2_{m + \lambda -1/2}\Big ((m + \lambda ) \varepsilon + (m + \lambda - 1) \omega \frac{e^2 a^2}{2 (pk)^2} (k p_{(\lambda )}) \Big ) + \cr & \displaystyle + (1 - 2 \lambda \cos \theta ) J^2_{m + \lambda +1/2}\Big ((m + \lambda ) \varepsilon + (m + \lambda + 1) \omega \frac{e^2 a^2}{2 (pk)^2} (k p_{(-\lambda )}) \Big ) \Big ). \label{Eq5.23}
\end{eqnarray}
\end{widetext}
\begin{figure*}
\center
\includegraphics[width=13.00cm, height=9.00cm]{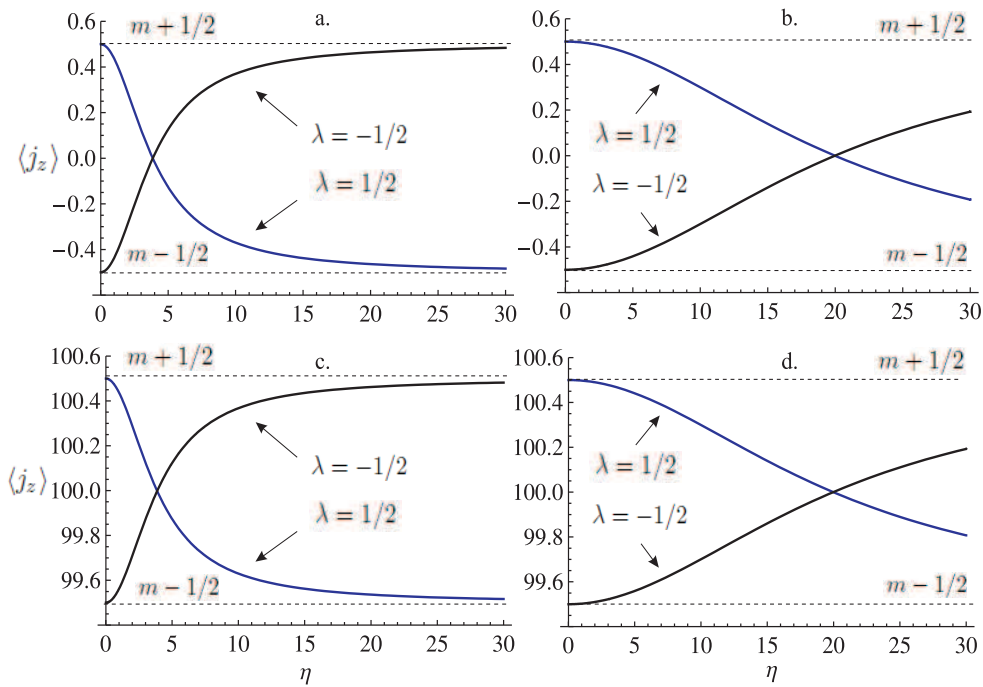}
\caption{\label{Fig1} The electron's effective total angular momentum (in units of $\hbar$) as a function of the laser wave's intensity. Upper panel (a, b): $m = 0, \kappa \rightarrow 0$, lower panel (c, d): $m = 100, \kappa = 50$ keV. Left panel (a, c): $\gamma = 2$, right panel (b, d): $\gamma = 10$. The angular momentum is shifted due to precession of the spin.}
\end{figure*}

It is clear that the intensity-induced shift of $\langle \hat{j}_z \rangle$ is negligible when $m \gg 1$, so the total angular momentum in this case is 
\begin{eqnarray}
& \displaystyle \langle \hat{j}_z \rangle \simeq m + \lambda \simeq m \nonumber
\end{eqnarray}
being the ``effective'' integral of motion. In Fig.\ref{Fig1} we represent this shift as a function of the laser intensity parameter $\eta$ in two cases: non-relativistic electron with the Lorentz-factor $\gamma = 2$ and the relativistic one with $\gamma = 10$. As can be seen in Fig.\ref{Fig1}, in the formal limit $\eta \gg 1$, which is known to correspond to a constant crossed field \cite{R, BLP}, the electron's effective helicity changes its sign (a non-radiative spin-flip due to precession) that can be significant for non-relativistic electrons with the small values of $m$ only. It can also be shown that the plots for $m \ne 0$ are almost independent of the $\kappa$'s value. Nevertheless, high-power lasers providing the values $\eta \gg 1$ already exist (such as, for example, Vulcan \cite{Vulcan}) and even more powerful facilities are under construction (such as ELI \cite{ELI}). It means that such an angular momentum's shift can, in principle, influence some quantum processes in the strong laser fields.

Finally, note that in the current model of the circularly polarized wave the time average in Eq.(\ref{Eq5.23}) was not performed. However, such an average, as easy to show, is equivalent to the integration over the azimuthal angle $\phi_{\mathcal R}$. In other words, the very same formulas for OAM and spin could be obtained a bit easier when initially performing the average over $\mathcal R^0$ that would result in the striking out the terms like $(\tilde{{\bm H}} {\bm \Sigma})$ and in the replacement of $(\tilde{{\bm H}} {\bm \Sigma}) (\tilde{{\bm E}} {\bm \alpha})$ with $\omega^2 a^2 [{\bm \alpha}\times {\bm \Sigma }]_z /2$ in Eq.(\ref{Eq5.20}). On the other hand, in the paraxial limit ($\kappa \rightarrow 0$) Eq.(\ref{Eq5.23}) for $ \langle \hat{j}_z \rangle$ transforms into Eq.(\ref{Eq5.11}) for $ \langle \bar {\hat{s}}_z \rangle$ multiplied by $\delta_{m, 0}$. This supports the claim that the time average used when deriving Eq.(\ref{Eq5.11}) is actually unnecessary for this model of the laser wave: we just omitted from the very beginning the terms, which would anyway vanish after the integration over the azimuthal angle.

\section{\label{Sect6} Discussion} 

Until now, only twisted electrons with the energies not higher than $300$ keV are experimentally realized, so the crucial question is how to accelerate them up to relativistic energies without losing the OAM. Since the direct experimental production of the relativistic twisted electrons seems to be unrealizable due to extremely small de Broglie wave length, the simplest way is to apply some external field for which the angular momentum's projection onto the direction of motion (z-axis) is conserved in time, at least, on average. Such a field is a plane wave, but it actually cannot accelerate particles effectively. In order to transmit some energy to the particle, it is preferable to have a non-zero z-component of electric field. We see two principal possibilities:

$\bullet$ One can use the focused laser beams whose TM-modes are known to have the non-zero z-components of electric field. It is also known that at least for the weakly focused (along z-axis) laser waves, the z-projection of the electron's angular momentum is an integral of motion \cite{N}. On the other hand, a tightly focused laser beam can acquire its intrinsic spin-orbit coupling that makes such a beam equivalent to a Bessel electromagnetic beam with the small (but non-zero!) value of $m$ (see, for example, Eq.(3) in \cite{F}). The trapping of charged particles in such beams can be effective, as has recently been discussed in \cite{Bagrov-07, Iwo}, though a conservation of the electron's OAM in this field is questionable for its not very high values. Nevertheless, this technique seems to be applicable for twisted electrons with $m \gg 1$. 

$\bullet$ Another possible way is to combine the longitudinal electric field, which will accelerate electrons, together with the co-propagating laser field and (or) the longitudinal magnetic field, which will trap and focus the electrons, by analogy with the usual electron gun. The conservation of the averaged angular momentum in such a scheme follows from the fact that both the electric field and the plane-wave field can always be effectively excluded from the equations of motion as well as from the Klein-Gordon equation \cite{Bagrov}. In other words, the problem of motion in the so-called ``combined'' field ($E_z + H_z + A$ with ${\bf k} = k {\bf e}_z$; see, for example, \cite{Bagrov, Bagrov-07, Bagrov-02}) can be reduced to the well-known problem of motion in a constant z-directed magnetic field. The key fact here is that the electron OAM's z-projection is an integral of motion for the constant magnetic field as for a free particle (see, for example, \cite{Bagrov-02}). It allows one to effectively apply the ``combined'' field configuration for accelerating non-relativistic twisted electrons.

For the sake of convenience, we demonstrate here the simplest example of the effective excluding the laser wave from the twisted states. Firstly, one could expand the initial state $|\psi \rangle$ in (\ref{Eq4.1}) over the $|q\rangle$-states instead of $|p\rangle$. For this purpose, one should rewrite the Volkov solution in terms of the quasi-momentum $q$. For the circularly polarized laser wave from Eq.(\ref{Eq4.6}), after the integration by parts one can present the action in the following way:
\begin{eqnarray}
& \displaystyle S = -(pr) - \frac{e}{(pk)} \int d \varphi\ \Big ((pA) - \frac{e}{2} A^2 \Big ) = - (q \mathcal R).
\label{Eq5.30}
\end{eqnarray}
This action has a form of the one for a free electron with replacements: $p \rightarrow q, r\rightarrow \mathcal R$ (compare with \cite{Bagrov}). In addition, instead of the free-electron bispinor $u(p)$, which enters the Volkov solution, it is necessary to take the bispinor $u(q)$ obeying the equation $(\gamma q - m_e) u(q) = 0$. As a result, we find the required $| q \rangle$-states:
\begin{eqnarray}
& \displaystyle \psi_q (r) = N \Big ( 1 + \frac{e}{2(qk)} (\gamma k) (\gamma A)\Big ) u(q) e^{-i (q \mathcal R)},
\label{Eq5.31}
\end{eqnarray}
Formally, this expression differs from the plane-wave state of the free electron only in the spin term and from the Volkov solution it does only in the replacement $u(p) \rightarrow u(q)$. Expanding now the twisted state over these $| q \rangle$-states, as in (\ref{Eq4.1}), we obtain for $\psi (r)$ the formula, which differs from Eq.(\ref{Eq4.8}) only in that the bispinors $u^{(\pm 1/2)}$ now depend on the polar angle of the vector ${\bm q}$ instead of the one of ${\bm p}$, and also on $q^0$ instead of $\varepsilon$\footnote{Hence, the helicity states are defined as $({\bm \sigma} {\bm n}) w = 2 \lambda w,\ {\bm n} = {\bm q}/q$.}.

Thus, the field of a plane wave can be effectively excluded from the Volkov solution of the Klein-Gordon equation. The same trick can be applied to the longitudinal electric field as well \cite{Bagrov}. It means the electron's effective angular momentum is conserved in the ``combined'' field (neglecting the spin's shift) that allows one to accelerate non-relativistic twisted electrons without a loss of OAM. 

On the other hand, this conservation is justified only when neglecting the radiation of photons, which can take the electron's OAM away. In order to estimate the OAM radiation losses, it is necessary to calculate the amplitude of the non-linear Compton effect with a twisted electron. For the weak laser fields with $\eta \ll 1$ and the big values of OAM, $m \gg 1$, one can use the ``bare'' QED with the scalar twisted particles, as was demonstrated in \cite{Ivanov, I-S}.

Let us now discuss some new effects the electron's OAM can bring into high-energy physics. Though an effective reduction of the spin (or even inversion of the one) in a strong laser field seems to be negligible for the twisted electron with $m \gg 1$, it can bring about some changes in the quantum cascade processes in the strong fields of the modern lasers with $\eta \gtrsim 1$. Indeed, when propagating through a beam of the strong laser the plane-wave electron with $m = 0$ is known to produce an electron-positron avalanche via the nonlinear Compton scattering and the Breit-Wheeler pair production processes \cite{Lim}. The parameters of such an avalanche were shown to be very sensitive to polarization of the laser beams \cite{Bulanov}. It is clear that changes in the effective polarization state of the electrons and positrons (due to precession of their spins in a strong field) would lead to some variations in the scattering cross section that can, therefore, influence the avalanche-like process. More detailed calculations are required to see how strong this effect is.

On the other hand, the shift of the electron's helicity just illustrates the obvious fact that the helicity is no longer an integral of motion for the electron in the plane-wave field. One can suppose that the choice of another spin quantum number, whose spin operator commutes with the Dirac Hamiltonian, would be more effective, since there will be no spin shift whatsoever in this case. Such spin operators, which are integrals of motion, are given, for example, in \cite{Bagrov-02}, however, their physical interpretation stays unclear.

Another possible effect is an increase of the electron's magnetic moment due to OAM that can lead, in particular, to the substantial rise in the energy radiated by this moment in some external field. 
The question is far from being academic. Indeed, it is well known that the anomalous magnetic moment of electron, which is proportional to the fine structure constant and is almost $10^3$ times smaller than the Bohr magneton, changes the radiation properties dramatically. In particular, it reduces the time of radiative polarization of the electron (positron) beams due to synchrotron radiation \cite{Bord}. The similar ``positive'' effect may take place for the radiative polarization in a helical undulator \cite{H}. On the contrary, the OAM-induced magnetic moment of electron is proportional to the OAM itself becoming approximately $10^2$ times lager for a twisted electron with $m \sim 100 \hbar$ experimental production of which seems to be not a problem, at least, for energies of $200 - 300$ keV. Since in the lowest order of perturbation theory the radiated energy $W \propto \mu^2$, for a twisted electron with $m \sim 100 \hbar$ the radiation power increases in $\sim 10^4$ times. This estimate becomes even more optimistic when considering the coherent regime of emission of the electron beam, as was discussed for the plane-wave electrons by Gover in \cite{Gover}. It seems that the most suitable process for observation of this effect is exactly the head-on collision of twisted electrons with the laser pulses, since the OAM stays almost stable in this case (in contrast to the synchrotron radiation scheme, in which the ``longitudinal orbital polarization'' will turn into the transverse one). Of course all the calculations of the radiation by magnetic moments can be performed within the framework of classical electrodynamics, as it is often done for plane-wave electrons (see e.g. \cite{Bord}), that allows one to consider, in addition, such processes as Vavilov-Cherenkov radiation, transition radiation, etc.

Finally, another possible application of twisted electrons, positrons and other particles with $m \gg 1$ is worth noting. It is well-known that the polarization states of particles are of the great importance for a number of processes to be studied at the future colliders such as the Compact Linear Collider and the International Linear Collider (see, for example, \cite{Zimm, Bailey}). Since the value of the particle's OAM can be, in principal, arbitrary large, the influence of the OAM degree of freedom on the different quantum processes may turn out to be even more powerful than that of the spin. While a production of the highly polarized positron beams for the next generation colliders seems to be a serious problem, it can be reasonable to consider the alternative schemes with OAM-polarized twisted beams. 

The production of twisted electrons in this case can be realized by analogy with the conventional source of the longitudinally polarized electrons based on the laser-driven photocathode (see e.g. \cite{C}). Namely, one can expect that if the initial laser photons carry OAM, the resultant electrons will be twisted as well that can also be used as another way for creating twisted electrons.

\section{\label{Sect7} Conclusion}

In most cases, the plane-wave states of particles are used in calculations of high-energy physics. Nevertheless, the quantum processes where such states are inapplicable are well-known (see, for example, \cite{Impact}). The massive particles carrying orbital angular momentum, which may be called the twisted particles by analogy with the photons \cite{Tw}, represent one of the simplest examples of the non-plane-wave states preserving the azimuthal symmetry. The experimental production of these particles (namely, electrons with $m$ up to $\sim 100 \hbar$) by several groups at the almost same time opens up possibilities for studying a number of new physical effects, the simplest of which is an increase of the electron's magnetic moment. 

On the other hand, there exist some other non-plane wave states being explored nowadays in quantum optics including the azimuthally non-symmetric Airy electron beams recently obtained in \cite{Bloch}. An inclusion of all these states into research area of the high energy physics requires, along with the other topics, some adaptation of the S-matrix formalism developed mostly for the plane wave states.

The OAM, as a new degree of freedom, may turn out to be highly important for a number of processes in atomic physics, nuclear physics, particle physics, astrophysics, etc., since the values of OAM can be much higher than those of spin. In particular, the electron's twisted states in external plane electromagnetic wave, which we have presented in this paper, allow one to calculate such quantum processes as the non-linear Compton scattering and the Breit-Wheeler pair production. On the other hand, the exact solutions of relativistic wave equations describing the twisted states enable the study of motion of these particles in external fields that can, consequently, suggest a way for accelerating non-relativistic electrons with OAM. As we have demonstrated, the special combination of fields makes the acceleration of twisted electrons possible.

\acknowledgments
Author is grateful to V.G. Bagrov, P.O. Kazinski, A.P. Potylitsyn, N.F. Shul'ga and, especially, to Igor Ivanov for stimulating criticism and fruitful discussions. This work is partially supported by the Russian Ministry for Education and Science within the program ``Nauka'', under the contracts Nos. $\Pi$1199, 14.B37.21.0911, 14.B37.21.1298 and by the Russian Foundation for Basic Research under the contract No.12-02-31071-mol\_a.

\end{document}